# A Systematic Post-Processing Approach for Quantitative $T_{1\rho}$ Imaging of Knee Articular Cartilage

Junru Zhong*, Yongcheng Yao*#, Fan Xiao, Tim-Yun Michael Ong, Ki-Wai Kevin Ho, Siyue Li#, Chaoxing Huang, Queenie Chan, James F. Griffith, Weitian Chen

*Authors share the same contribution. # Work was partially done when Y. Yao and S. Li was with CUHK.

J. Zhong, C. Huang, J. Griffith, and W. Chen are affiliated with the CU Lab of AI in Radiology, Department of Imaging and Interventional Radiology, Faculty of Medicine, The Chinese University of Hong Kong.

Y. Yao is affiliated with the CU Lab of AI in Radiology, Department of Imaging and Interventional Radiology, Faculty of Medicine, The Chinese University of Hong Kong; and the School of Informatics, University of Edinburgh.

F. Xiao is affiliated with the Department of Radiology, Shanghai Sixth People's Hospital Affiliated to Shanghai Jiao Tong University School of Medicine.

M. Ong and K. Ho are affiliated with the Department of Orthopaedics & Traumatology, Faculty of Medicine, The Chinese University of Hong Kong.

S. Li is affiliated with the CU Lab of AI in Radiology, Department of Imaging and Interventional Radiology, Faculty of Medicine, The Chinese University of Hong Kong; and Magnetic Resonance Research Lab, Department of Radiological Science, UCLA.

Q. Chan is affiliated with Philips Healthcare.

**Word Count for Text:**

Main text 3185 words; abstract 237 words

Post-proc approach for knee $T_{1\rho}$ imaging

# Abstract

**Objective** To establish an automated pipeline for post-processing of quantitative spin-lattice relaxation time constant in the rotating frame ($T_{1\rho}$) imaging of knee articular cartilage.

**Design** The proposed post-processing pipeline commences with an image standardisation procedure, followed by deep learning-based segmentation to generate cartilage masks. The articular cartilage is then automatically parcellated into 20 subregions, where $T_{1\rho}$ quantification is performed. The proposed pipeline was retrospectively validated on a dataset comprising knee $T_{1\rho}$ images of 10 healthy volunteers and 30 patients with knee osteoarthritis. Three experiments were conducted, namely an assessment of segmentation model performance (using Dice similarity coefficients, DSCs); an evaluation of the impact of standardisation; and a test of $T_{1\rho}$ quantification accuracy (using paired *t*-tests; root-mean-square deviations, RMSDs; and coefficients of variance of RMSDs, $CV_{RMSD}$). Statistical significance was set as $p < 0.05$.

**Results** There was a substantial agreement between the subregional $T_{1\rho}$ quantification from the model-predicted masks and those from the manual segmentation labels. In patients, 17 of 20 subregions, and in healthy volunteers, 18 out of 20 subregions, demonstrated no significant difference between predicted and reference $T_{1\rho}$ quantifications. Average RMSDs were 0.79 ms for patients and 0.56 ms for healthy volunteers, while average $CV_{RMSD}$ were 1.97% and 1.38% for patients and healthy volunteers. Bland–Altman plots showed negligible bias across all subregions for patients and healthy volunteers.

**Conclusion** The proposed pipeline can perform automatic and reliable post-processing of quantitative $T_{1\rho}$ images of knee articular cartilage.

**Keywords**: $T_{1\rho}$ imaging, quantitative MRI, post-processing, deep learning

# Introduction

Quantitative magnetic resonance imaging (qMRI) mapping of cartilage is an emerging topic in musculoskeletal (MSK) imaging. It is a family of MRI techniques that allow quantitative measurement of cartilage properties, enabling the early detection of joint diseases such as knee osteoarthritis (OA) [1]. Among the methods applied, spin-lattice relaxation time constant in the rotating frame ($T_{1\rho}$) mapping [2] is sensitive to the proteoglycan content of human cartilage [3,4], which is a biomarker of its degeneration. There is a correlation between $T_{1\rho}$ values and knee OA: Wang et al. [5] showed a statistically significant difference between patients with varying severity of OA in terms of $T_{1\rho}$ values in the meniscus and articular cartilage, and Prasad et al. [6] found that $T_{1\rho}$ values increased as the severity of OA increased over a 2-year period. In addition, $T_{1\rho}$ values have also been used to monitor cartilage recovery after surgery [7,8]. While $T_{1\rho}$ assessment is valuable for various MSK applications, it is not widely used in clinical practice.

Deep learning methods exhibit good performance in knee cartilage segmentation [9]. However, most deep learning methods have been developed for anatomical MRI, and few studies have examined segmentation in qMRI. Norman et al. reported an approach that leveraged a two-dimensional U-Net to automate segmentation in knee $T_{1\rho}$ MRI [10]. The authors trained U-Nets for segmenting articular cartilage and meniscus on an internal $T_{1\rho}$-weighted magnetisation-prepared angle-modulated-partitioned *k*-space-spoiled gradient-echo snapshots (MAPSS) sequence dataset and a three-dimensional (3D) double echo steady state (DESS) dataset from the Osteoarthritis Initiative project [11]. Similarly, Desai et al. [12] created a dataset with a quantitative DESS sequence for $T_2$ mapping and manual labelling of cartilage and meniscus and developed a corresponding segmentation method. Both studies have demonstrated that deep learning methods can achieve satisfactory segmentation performance and quantification correlations in qMRI of the knee.

Subregion analysis is essential for post-processing of qMRI of the knee, with most studies relying on manual region-of-interest (ROI) selection [1,5,8,13]. However, few studies have investigated automatic cartilage subregion parcellation of qMRI of the knee. Development of a reliable technique for automatic subregion parcellation of quantitative $T_{1\rho}$ images of knee articular cartilage would be very helpful clinically.

In this study, we propose a pipeline for automatic post-processing of quantitative $T_{1\rho}$ imaging of knee cartilage, thereby facilitating the use of this promising MRI technique in routine clinical applications. The key contributions of this study are as follows. (i) Build an automatic knee qMRI post-processing pipeline with state-of-the-art deep learning methods. (ii) Implement a rule-based and sequence-independent parcellation method to facilitate regional quantification of cartilage. (iii) Demonstrate the consistent and robust performance of the proposed algorithms in evaluation of simulated and in-house $T_{1\rho}$-weighted MRI data.

# Materials and Methods

## Participants, MRI Acquisition, and Data Processing

The study had been approved by our institutional ethics board, The Joint CUHK-NTEC CREC. The study procedures followed were in accordance with the ethical standards of the responsible committee on human experimentation (institutional and national) and with the Helsinki Declaration of 1975, as revised in 2000. The participants comprised 30 OA patients and 10 healthy volunteers. All participants provided informed consent and underwent MRI examinations of one knee on a Philips Achieva TX 3.0 T scanner (Philips Healthcare, Best, Netherlands) from June to October 2022. $T_{1\rho}$ MR images ($I$) were collected using a magnetisation-prepared 3D turbo spin echo (TSE) acquisition [14–16]. The details of MRI parameters are given in Table 1. Segmentation labels ($S$) were prepared on femoral cartilage (FC), medial tibial cartilage (MTC), lateral tibial cartilage (LTC), and patellar cartilage (PC) for all samples under the supervision of a radiologist (F. Xiao) with 8 years' experience of knee MRI reporting.

## Proposed Knee $T_{1\rho}$ Post-Processing Pipeline

Our proposed knee $T_{1\rho}$ MRI post-processing pipeline consists of four modules: image standardisation, cartilage segmentation, subregion parcellation, and $T_{1\rho}$ quantification modules. Figure 1a shows an overview of the pipeline.

## Image Standardisation

The pipeline starts with an image standardisation procedure to improve the robustness of our proposed pipeline and alleviate the effect of variations in MRI examination protocols, prescriptions, and subject positioning. Our rule-based parcellation algorithm is defined in the 'RAS+' coordinate system, *i.e.*, the positive directions of the data array coordinates are aligned to the right, anterior, and superior directions in the real-world coordinate system. This step is necessary to ensure proper parcellation.

The proposed image standardisation involves two steps. First, the image $I$ is aligned to an 'RAS+' coordinate system, *i.e.*, the positive direction of the data array coordinates is aligned to the right, anterior, and superior directions in the real-world coordinate system. Next, the image $I$ is further aligned to a chosen reference image via rigid-body registration. In addition, the corresponding cartilage masks $\hat{S}$ are registered to the same coordinates using the transform matrix. Subsequent steps, *i.e.*, subregion parcellation and $T_{1\rho}$ quantification, are based on standardised images. Figure 1b shows the standardisation process.

## Cartilage Segmentation

Cartilage segmentation is the foundation of the pipeline. Segmentation provides the masks of the cartilage compartments, allowing further localisation of the cartilage subregions. We exploit an advanced but easy-to-use segmentation model, nnU-Net, for cartilage segmentation in the pipeline. nnU-Net [17] is a state-of-the-art DL-based segmentation method that has shown excellent performance when applied to different medical imaging datasets, such as knee cartilage segmentation [18,19].

In the current study, we employed five-fold cross-validation to train 3D nnU-Net models $F_{seg}$ on a dataset comprising a mixture of data samples from patients and healthy volunteers ($n = 40$). The models were trained using the mean image $\bar{I}$ of the four dynamic images $I$ and the manual segmentation mask $S$, where $\bar{I}$ was used to improve the contrast-to-noise ratio. We collected the prediction results from the five folds and formed the full prediction result $\hat{S}$ for further demonstration and evaluation.

## Cartilage Subregions Parcellation

We partitioned the femoral and tibial cartilages into 20 subregions on femoral and tibial cartilage masks using an algorithm adopted from section 3.7 of Yao et al.[18]. Early studies [20,21] about regional analysis of cartilage loss inspired the parcellation scheme, which is described below.

In this scheme, FC is partitioned into medial and lateral compartments (*i.e.* medial FC (MFC) and lateral FC (LFC)), each of which is further segmented into anterior, central, and posterior parts (*e.g.*, anterior MFC (aMFC), central MFC (cMFC), and posterior MFC (pMFC)). As we are interested in the early alteration of $T_{1\rho}$ magnitudes in weight-bearing regions, the central femoral areas are split into three subregions in the left-right direction, namely the exterior, central, and interior central FC (*e.g.*, exterior central MFC (ecMFC), central central MFC (ccMFC), and interior central MFC (icMFC)). The subregions of the MTC or LTC consist of an elliptical central ROI (*e.g.*, cMTC) and four peripheral ROIs, i.e., the anterior, posterior, exterior, and interior compartments (*e.g.*, anterior MTC (aMTC), posterior MTC (pMTC), exterior MTC (eMTC), and interior MTC (iMTC)). An example of parcellation is shown in Figure 2.

Our rule-based parcellation algorithm only accepts the cartilage masks from images fitting the 'RAS+' coordinate system. A prerequisite is for the patella to be parallel to the horizontal line in the axial view. If this prerequisite is not met, incorrect sub-regional masks are produced. The use of the proposed image standardisation algorithm means that all processed images satisfy this requirement.

## $T_{1\rho}$ Quantification

$$I(tsl) = I_0 \times \exp\left(\frac{-tsl}{T_{1\rho}}\right) + c$$

(1)

Equation (1) is a scalar formula that is typically used to fit $T_{1\rho}$ [22,23], where *tsl* is the time of the spin-lock; *I(tsl)* is the $T_{1\rho}$-weighted image acquired at *tsl*; $I_0$ is the image at *tsl* = 0; and *c* is a constant that can be zero or non-zero, depending on the pulse sequences and fitting methods. Various fitting methods have been proposed to obtain $T_{1\rho}$ maps using Equation (1) [22], including the recently developed deep-learning-based methods that require training data [24,25].

In the current study, the MRI protocol we used to obtain data collects four $T_{1\rho}$-weighted images. $T_{1\rho}$ is quantified pixel-wise by using a variant of the dichotomy method [26] to fit images to the relaxation model described in Equation (1). As shown in Figure 1c, we performed pixel-wise $T_{1\rho}$ fitting within the cartilage subregions. The mean and other statistics of $T_{1\rho}$ values within each subregion after the parcellation are provided as the output of the proposed post-processing pipeline.

# Evaluation Experiments

We evaluated the proposed pipeline in three experiments. **Experiment 1** studied the automated segmentation of the pipeline; **Experiment 2** analysed the effect of standardisation on the rule-based parcellation; and **Experiment 3** evaluated the performance of the entire pipeline.

## Experiment 1: Automated segmentation

We assessed the performance of our method on the entire dataset ($n = 40$) using standard evaluation metrics such as the Dice similarity coefficient (DSC) and the average symmetric surface distance (ASSD). The definitions of DSC and ASSD are provided in section A of the Supplementary Material. Additionally, we conducted statistical analyses on the regional $T_{1\rho}$ quantification obtained from ground truth ($S$) and predicted ($\hat{S}$) segmentation. The detailed methods are given in the *Statistical Analysis* section.

## Experiment 2: Effectiveness of standardisation

This experiment evaluated the effect of standardisation on the parcellation of $T_{1\rho}$-weighted images with various shapes and orientation. Two studies were performed: (i) a controlled study in which the images and segmentation labels of two healthy volunteers were rotated manually, and (ii) a real-world study of all data samples to observe improvements in parcellation accuracy after standardisation. Figure 3 demonstrates the procedure of the study (horizontal axis) and the pipeline output (vertical axis) when rotating an example image along the superior-inferior axis. Selected examples from the real-world study are shown in the Results section.

## Experiment 3: Evaluation of $T_{1\rho}$ quantification

The final experiment evaluated the entire proposed pipeline presented in Figure 1a. $T_{1\rho}$-weighted images ($I$) were standardised, segmented, parcellated, and quantified to produce automated subregional mean $T_{1\rho}$ values for 20 cartilage subregions. This experiment includes 27 OA patients and 10 healthy volunteers, excluding three OA patients who had more than 50% full-thickness cartilage loss (full-thickness cartilage loss can be quantified using CartiMorph Toolbox [18]) in any of the FC, MTC, and LTC compartments. This exclusion is justified as $T_{1\rho}$ mapping is mainly used to provide compositional information on cartilage in early-stage OA [9]. Subregional $T_{1\rho}$ quantifications generated from the ground truth and predicted cartilage masks, denoted as $Q$ and $\hat{Q}$, were statistically compared.

## Statistical Analysis

**Experiments 1** and **3** used paired-sample statistical tests, calculated root-mean-square deviations (RMSDs) and coefficients of variance of RMSD ($CV_{RMSD}$) to assess $T_{1\rho}$ quantifications at regional and subregional levels. Paired-sample t-tests (used in normally distributed data) or Wilcoxon tests (used in non-normally distributed data) were used to determine the differences in $T_{1\rho}$ quantifications, with a a non-significant result being expected. Normality (Shapiro-Wilk) tests were applied to the $T_{1\rho}$ quantification from each ROI to determine which paired-sample tests to use. We set the type-I ($\alpha$) and type-II ($\beta$) errors to 0.05 and 0.2, respectively, and the minimum sample size of 10 samples to ensure adequate statistical power for detecting a 10% difference (equivalent to 4 ms in our dataset) in $T_{1\rho}$ quantification between $Q$ and $\hat{Q}$.

RMSD and $CV_{RMSD}$ values are commonly used metrics in repeatability studies [15,27]. We applied them to directly evaluate the $T_{1\rho}$ quantification errors against the ground truth (**Experiment 1**) or reference (**Experiment 3**). In **Experiment 3**, Bland-Altman plots were employed to examine the performance of $q$-$\hat{q}$ pairs of each participant. The definitions of these metrics are given in section A.2 of the Supplementary Material.

## Implementation

We implemented the pipeline in MATLAB (version R2023b, MathWorks, Inc., Natick, MA, USA), nnU-Net (version 2.2) [17], PyTorch (version 2.1) [28], and Python (version 3.10). ITK-SNAP (version 4.0) [29] was used to prepare the ground-truth cartilage masks and prepare data in **Experiment 2**. We used 3D Slicer (version 5.6.1) [30] to visualise cartilage (subregion) masks. Statistical analysis complied with SciPy (version 1.11) [31], MedCalc (version 20.100, MedCalc Software Ltd, Ostend, Belgium) and Pingouin (version 0.5.3) [32]. All neural network training and validation were performed on an NVIDIA Quadro RTX 5000 GPU (Santa Clara, CA, USA).

# Results

## Demographics

Table 2 shows the demographics of the participants in the three experiments. A data group of 10 healthy volunteers (average age = 24.90 years, body mass index (BMI) = 22.875 kg/m$^2$, five men) was used in all three experiments. In addition, a data group of 30 patients was used in **Experiment 1** (average age = 67.63 years, BMI = 26.00 kg/m$^2$, nine men), and a data group of 27 patients (average age = 68.04 years, BMI = 25.44 kg/m$^2$, nine men) was used in **Experiment 3**.

## Experiment 1: Automated segmentation

The *Segmentation Performance* section of Table 3 shows that the automated segmentation achieved satisfactory performance on the FC, MTC, and LTC compartments in the patient and healthy volunteer groups in terms of segmentation metrics such as DSC and ASSD.

Regarding the $T_{1\rho}$ quantification, all three cartilage regions (FC, MTC, and LTC) in both the patient and healthy participant groups were not significantly different according to paired *t*-tests. Among the three regions, the average RMSDs for the patient and healthy volunteer groups were 0.34 ms and 0.26 ms, respectively, and the greatest RMSD was 0.47 ms, which was in the MTC patient group. Similarly, among the four regions, the average $CV_{RMSD}$ values of the patient and healthy volunteer groups were 0.88% and 0.68%, respectively, and the greatest $CV_{RMSD}$ value was 1.22%, which was in the MTC patient group. Detailed results are given in the *Regional Mean $T_{1\rho}$* section of Table 3. Normality test results are given in section B.1 of the Supplementary Material.

## Experiment 2: Effectiveness of standardisation

Figure 4a shows representative parcellation results from the controlled study, in which the images were manually rotated (complete results are given in section B.2 of the Supplementary Material). The column entitled 'Original' shows the accurate cartilage subregional masks from the parcellation algorithm without any rotation added. The column entitled 'Rotated' shows the incorrect cartilage parcellation resulting from large-degree rotations. The column entitled 'Standardised' shows the effectiveness of the proposed standardisation, *i.e.*, that a tilted image can be adjusted to ensure accurate partitioning of cartilage. Note that although only 3D visualisations are presented, the standardisation process involves MRI and cartilage masks.

The proposed standardisation improved the accuracy of cartilage parcellation for all data samples in our real-world study. Figure 4b shows representative results from three patients. A significant improvement can be observed in the boundary between aLFC (anterior lateral FC) and aMFC (anterior MFC), as shown in the axial view and 3D visualisation. The six subregions in central FC could be accurately delineated even with full-thickness cartilage loss (as seen in patient #068).

## Experiment 3: Evaluation of $T_{1\rho}$ quantification

In this experiment, we evaluated the performance of the entire proposed pipeline. Data from 9 subregions in the patients and 2 subregions in the healthy volunteers did not pass the normality test, necessitating the application of Wilcoxon tests for these datasets. Conversely, data from remaining subregions in both patient and healthy volunteer groups passed the normality test, thereby warranting the use of paired-sample *t*-tests for evaluation. etailed results of normality tests are given in section B.1 in the Supplementary Material.

Table 4 shows the results of this experiment. Among all subregions, iMTC (interior MTC), eLTC (exterior LTC), cLTC (central LTC) in the patient group and aLFC (anterior lateral FC), aLTC (anterior LTC) in the healthy volunteer group showed significant differences ($p < 0.05$) between $Q$ and $\hat{Q}$. The average RMSDs of all subregions from the patients and healthy volunteers were 0.79 *ms* and 0.56 *ms*, respectively, with the maximum RMSD of 1.73 *ms* in ccMFC (central central medial FC) in patients. The average $CV_{RMSD}$ values of all subregions from the patients and healthy volunteers were 1.97% and 1.38%, respectively, with the maximum $CV_{RMSD}$ of 4.58%, also in ccMFC of patients. Figure 5 and Figure 6 show the Bland-Altman plots of each $q$-$\hat{q}$ pair on patients and healthy volunteers, respectively. The bias was almost negligible for all subgroups in the patient and healthy volunteer groups. In addition, the limits of agreement (dashed lines) for most of these subgroups fell within ±10 ms in the patients and ±5 ms in the healthy volunteers.

# Discussion and Conclusion

$T_{1\rho}$ MRI holds promise for the early detection of knee OA [9]. However, the lack of an automated post-processing method hinders the adoption of $T_{1\rho}$ MRI in clinical routines. To solve this problem, we proposed a systematic post-processing pipeline for knee $T_{1\rho}$ MRI that utilises DL and image processing techniques to achieve automatic $T_{1\rho}$ quantification in knee cartilage subregions. We evaluated this proposed method on datasets collected using $T_{1\rho}$-prepared 3D TSE acquisitions [14–16]. As TSE is a staple pulse sequence in clinical knee MRI, such an acquisition technique has the potential to provide simultaneous clinical anatomical imaging and $T_{1\rho}$-based qMRI of the knee.

Our proposed post-processing pipeline can be extended to other $T_{1\rho}$ imaging methods, such as 3D MAPSS [33], as the image standardisation, parcellation, and $T_{1\rho}$ quantification are compatible with other acquisition techniques and do not depend on a specific pulse sequence. However, the DL-based segmentation we applied requires training data from particular pulse sequences. The need for a method for $T_{1\rho}$ quantification of subregions of knee cartilage has been highlighted [5,34,35]. In contrast to previous work [11,13], we move one step forward from cartilage segmentation to implement rule-based parcellation with standardisation. With the parcellation, our proposed pipeline can support a variety of clinical studies on $T_{1\rho}$ MRI of the knee.

We observed that although the segmentation performance in this study was poor compared with that in studies[10], there were statistically nonsignificant differences and minimal RMSD and $CV_{RMSD}$ values associated with $T_{1\rho}$ quantification in most subregions. This finding suggests that regional $T_{1\rho}$ quantification does not require accurate cartilage segmentation, possibly because $T_{1\rho}$ values have a relatively uniform distribution within each subregion. The average $T_{1\rho}$ value within the subregions does not deviate from the reference if the subregions after segmentation are within the cartilage region. Thus, even if segmentation cannot delineate the cartilage boundary accurately, reliable $T_{1\rho}$ quantification can nevertheless be obtained within subregions. This absence of a requirement for segmentation accuracy is a feature of quantitative $T_{1\rho}$ imaging that distinguishes it from cartilage quantification based on anatomical imaging [18]. Thus, further investigations are warranted to incorporate this feature into loss function to create a label-efficient segmentation, as this would reduce the cost of using this pipeline.

This study had some limitations. First, we used only a small and single-centre dataset collected by one protocol, and this may have impacted the performance of the proposed pipeline. Further studies are needed to evaluate the proposed pipeline in other centres and in different $T_{1\rho}$-based MR imaging techniques. Second, we implemented an automated parcellation using rules inspired by previous studies [20,21], which were designed without inclusion of patellar cartilage. Thus, the implementation of an algorithm with parcellation rules that includes patellar cartilage is worthy of further investigation. Last but not least, implementation of a graphical user interface is necessary for spreading the proposed pipeline to hospitals which can be achieved by extending existing work from Yao et al.[18].

In summary, we present a systematic post-processing pipeline for $T_{1\rho}$ imaging of knee cartilage, which we validated using both simulated and real-world data. The results demonstrate the pipeline's effectiveness for automated $T_{1\rho}$ quantification in subregions of knee cartilage. This automated post-processing pipeline can facilitate the application of quantitative $T_{1\rho}$ MRI of knee cartilage in routine clinical practice, thereby offering new direction for the diagnosis, prognosis, and treatment monitoring of knee osteoarthritis. The pipeline also has the potential to be applied in other knee qMRI techniques.

# Acknowledgements

We would like to acknowledge Chi Yin Ben Choi and Cheuk Nam Cherry Cheng for their assistance in patient recruitment and MRI exams, Tsz Shing Adam Kwong for his assistance in data processing, and Zongyou Cai for his advice in statistical analyses. The research was conducted in part at CUHK DIIR MRI Facility, which is jointly funded by Kai Chong Tong, HKSAR Research Matching Grant Scheme and the Department of Imaging and Interventional Radiology, The Chinese University of Hong Kong.

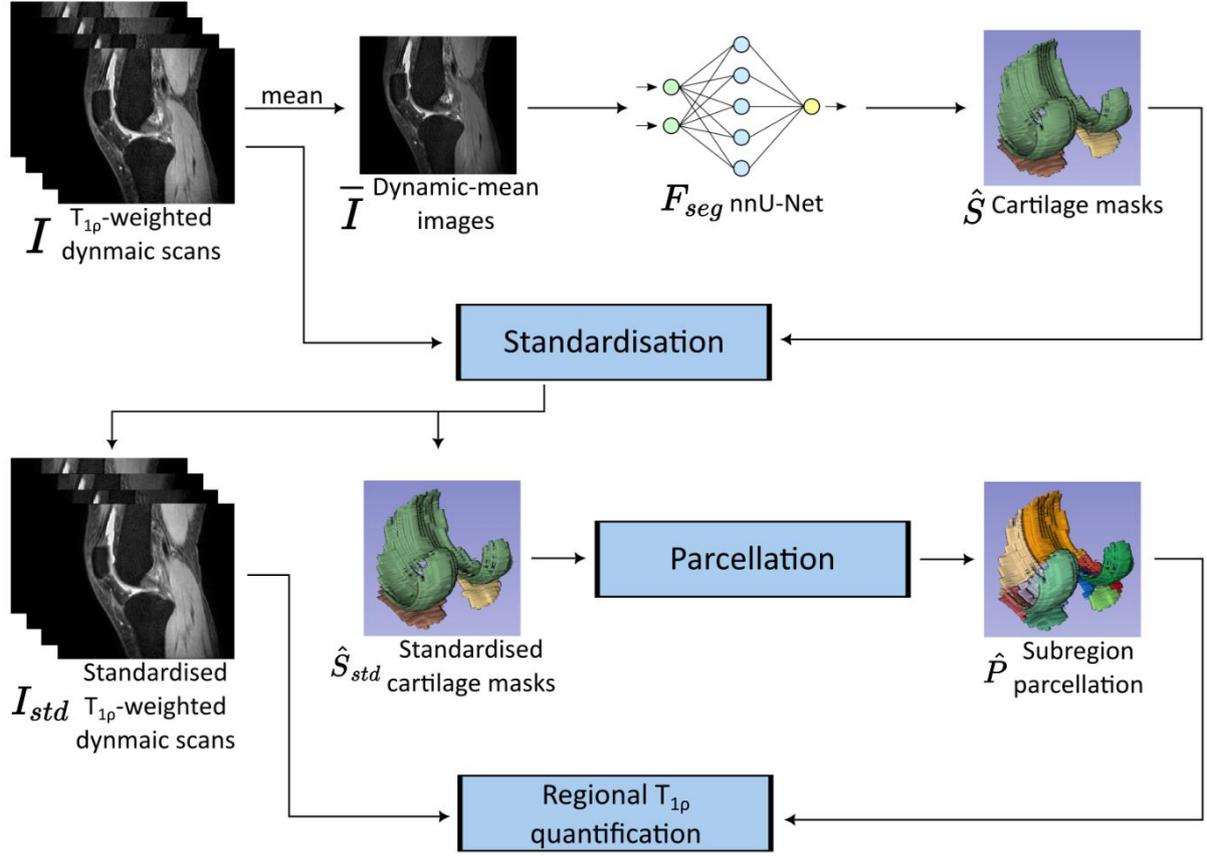

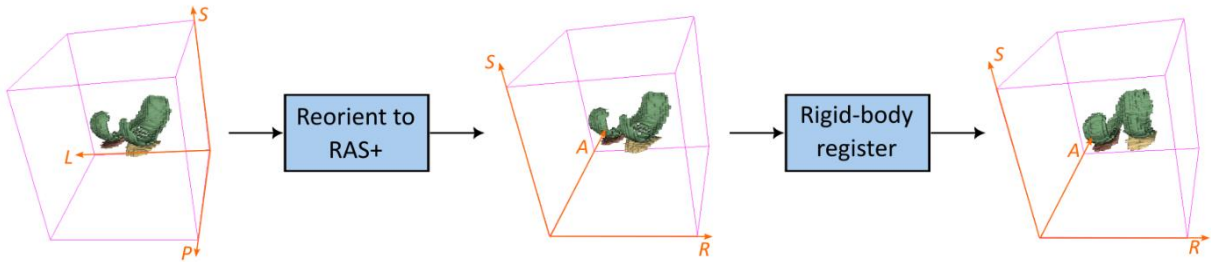

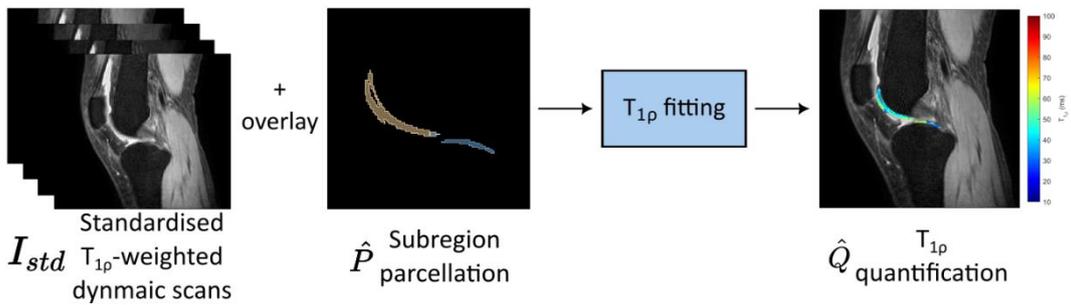

Figure 1. a. Overview of the proposed spin-lattice relaxation time constant in the rotating frame ($T_{1\rho}$) post-processing pipeline. b. The standardisation module. c. The regional $T_{1\rho}$ quantification. Note. Example images of a healthy participant (V014, right knee, female, 22 years, BMI = 20.76 kg/m$^2$) were processed. L = left, S = superior, P = posterior, A = anterior, R = right, BMI = body mass index.

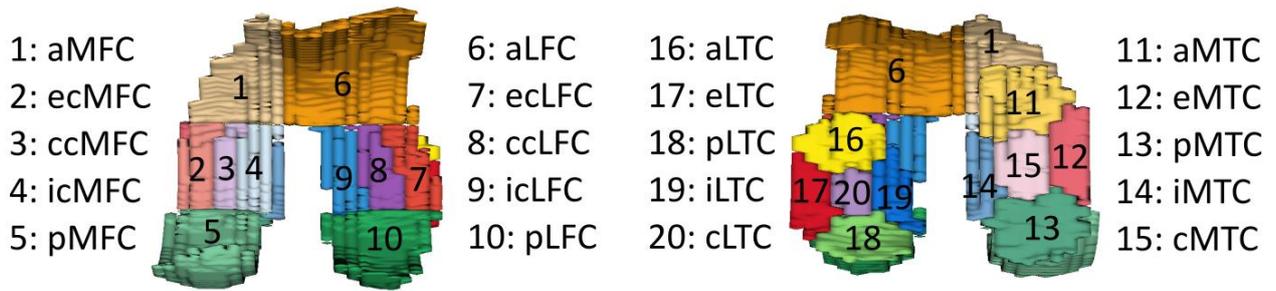

Figure 2. Parcellated cartilage subregions. Examples images of a healthy participant (V014, right knee, female, 22 y, 20.76 kg/m$^2$) were processed. Note. a = anterior, p = posterior, e = exterior, i = interior, c = central, ic = interior-central, cc = central-central, ec = exterior-central, MFC = medial femoral cartilage, LFC = lateral femoral cartilage, MTC = medial tibial cartilage, LTC = lateral femoral cartilage.

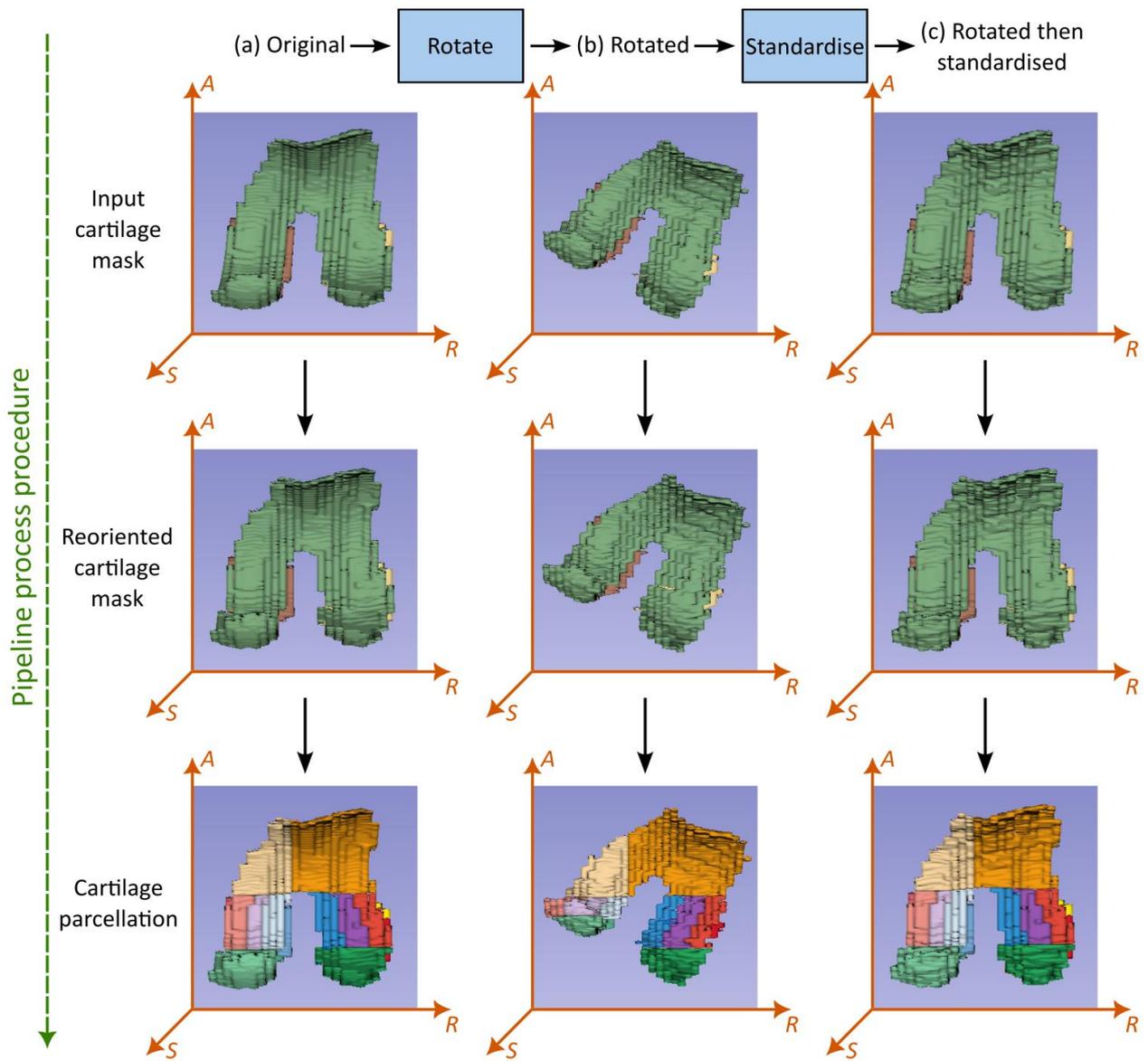

Figure 3. Process of the controlled study of **Experiment 2**. Example images of a healthy participant (V014, right knee, female, 22 years, BMI = 20.76 kg/m$^2$) were processed. Note: S = superior, A = anterior, R = right, BMI = body mass index.

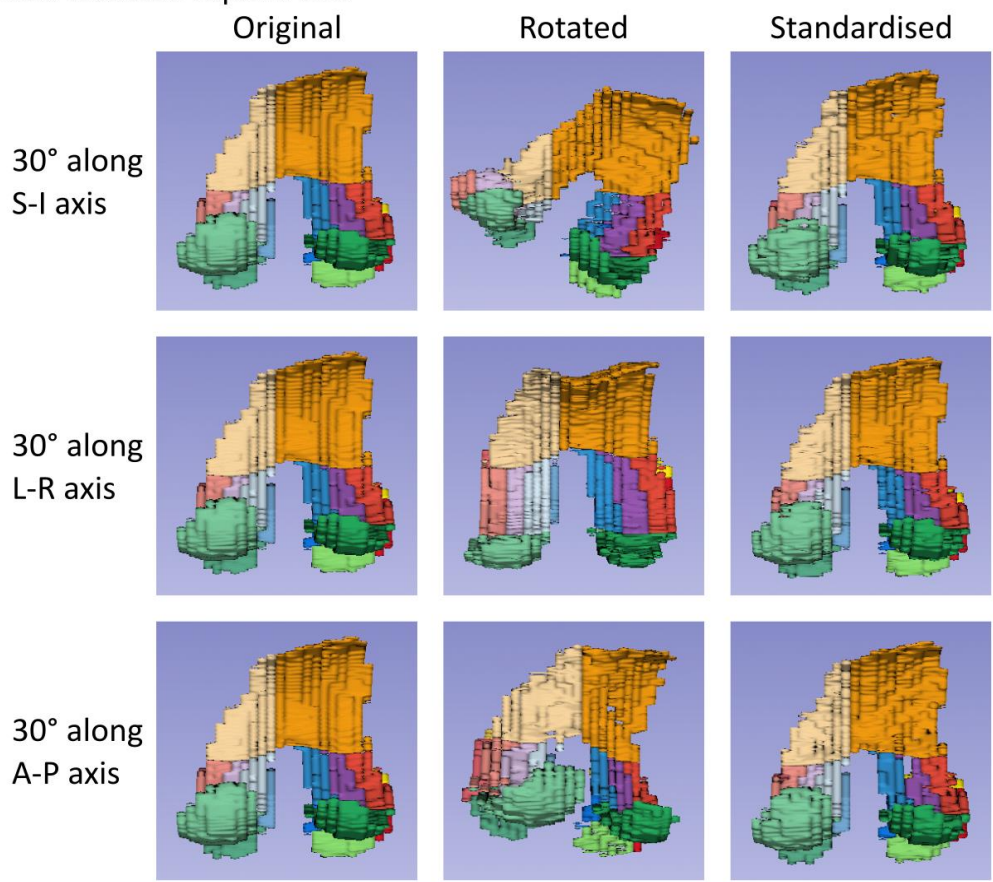

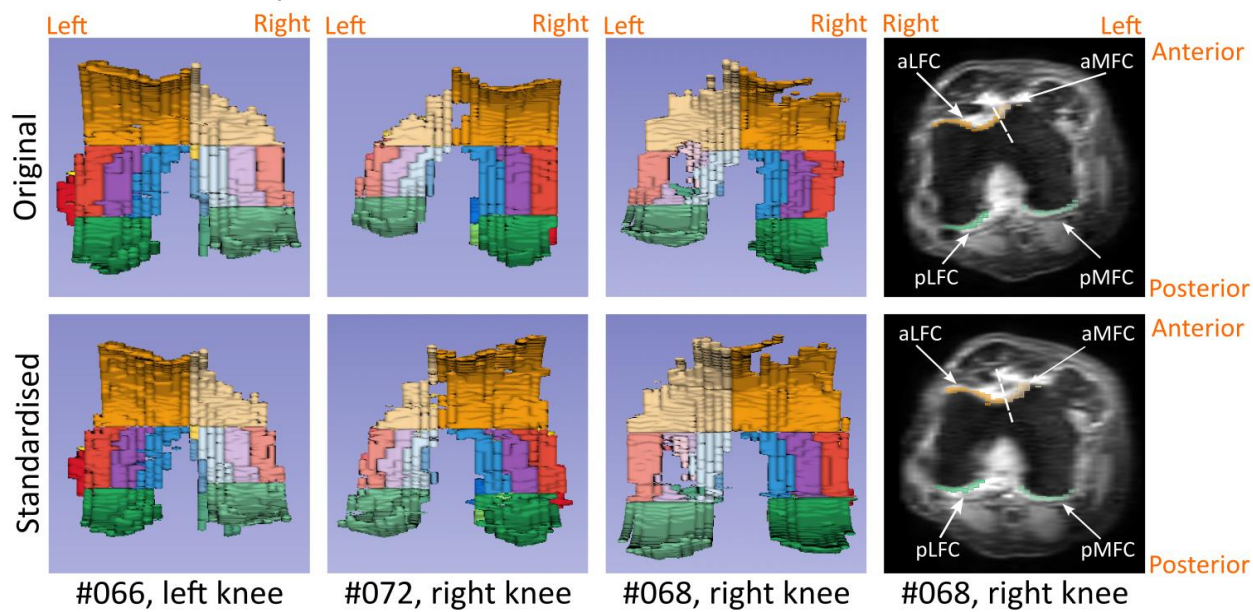

Figure 4. a. Controlled rotation experiment on a healthy participant (V014, right knee, female, 22 y, BMI = 20.76 kg/m$^2$). All images of the three dimensionally rendered masks were taken from a unified camera position. b. Standardisation examples from our patient data. Patient #066, left knee, female, 61 y, BMI = 26.98 kg/m$^2$, patient #072, right knee, male, 75 y, BMI = 28.10 kg/m$^2$, patient #068, right knee, male, 76 y, BMI = 27.08 kg/m$^2$. Note. BMI = body mass index.

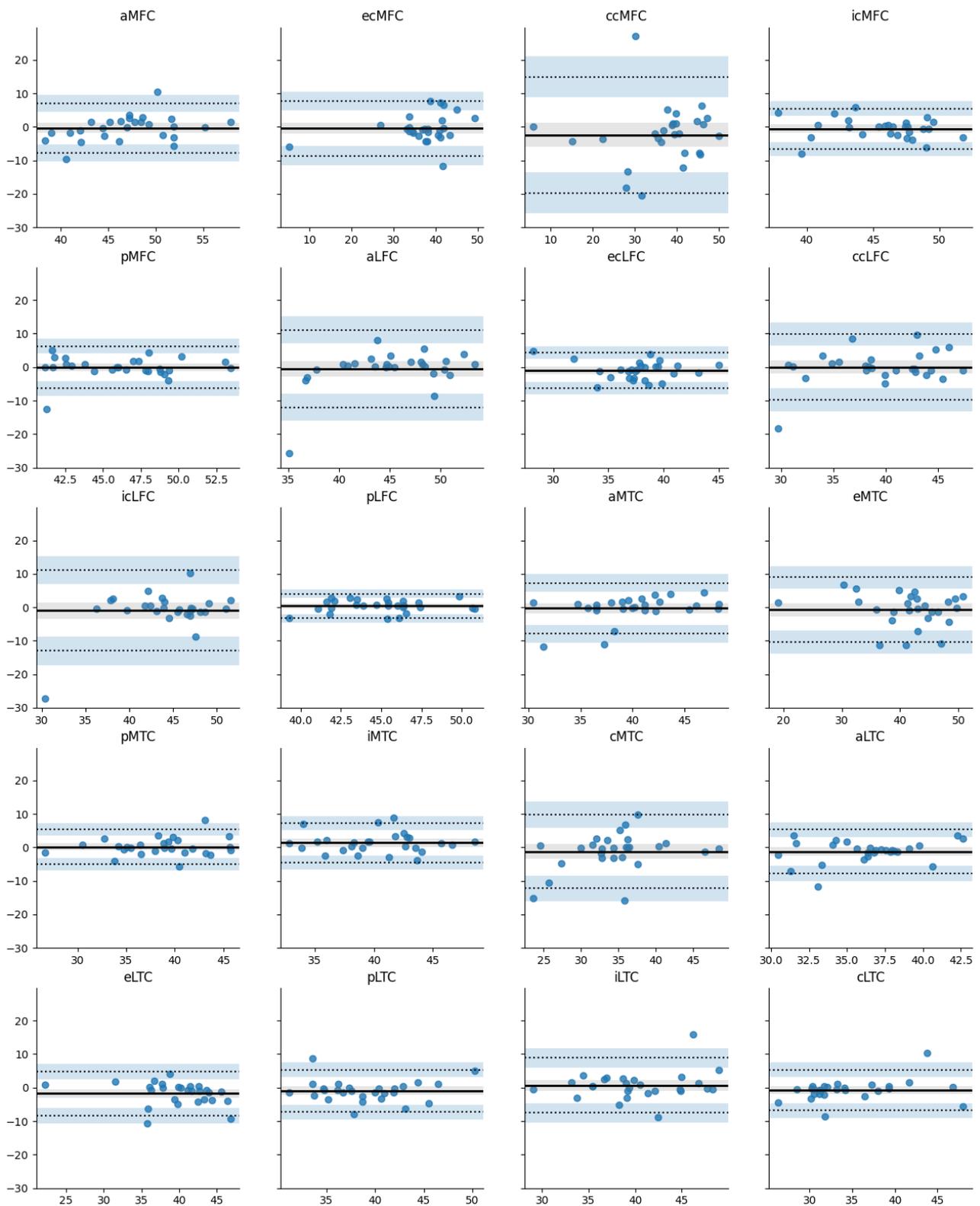

Figure 5. Bland-Altman plots for patients. The horizontal axes are the average of $q$-$\hat{q}$ pairs, whose ranges vary among subregions. The vertical axes are the differences between $q$-$\hat{q}$ pairs, whose ranges are unified as [-30, 20]. The unit of the numbers is *ms*. Note. a = anterior, p = posterior, e = exterior, i = interior, c = central, ic = interior central, cc = central central, ec = exterior central, MFC = medial femoral cartilage, LFC = lateral femoral cartilage, MTC = medial tibial cartilage, LTC = lateral femoral cartilage.

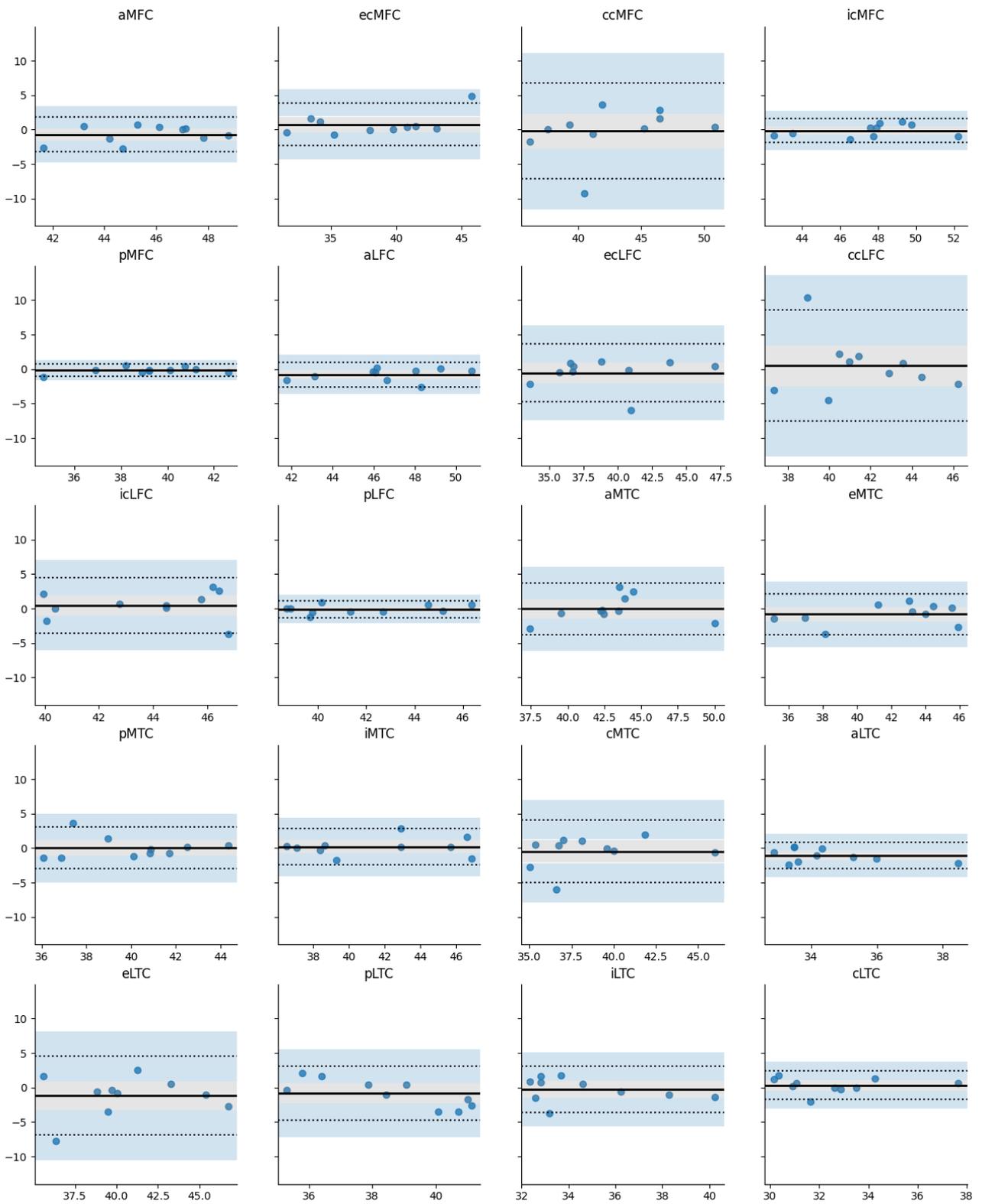

Figure 6. Bland-Altman plots for healthy volunteers. The horizontal axes are the average of q-$\hat{q}$ pairs, whose ranges vary among subregions. The vertical axes are the differences between q-$\hat{q}$ pairs, whose ranges are unified as [-10, 10]. The unit of the numbers is *ms*. Note. a = anterior, p = posterior, e = exterior, i = interior, c = central, ic = interior central, cc = central central, ec = exterior central, MFC = medial femoral cartilage, LFC = lateral femoral cartilage, MTC = medial tibial cartilage, LTC = lateral femoral cartilage.

**Table 1. Magnetic Resonance Imaging Acquisition Parameters**

| Item | Parameter |
|---|---|
| Plane | Sagittal |
| Fat suppression | SPAIR |
| No. of slices | 44 |
| Field of view (mm$^3$) | 160×160×132 |
| TE/TR (ms/ms) | 31/2000 |
| Resolution (mm$^3$) | 0.8×1×3 |
| Spin-lock Frequency (Hz) | 300 |
| Spin-lock Time (ms, in scanning order) | 0/50/30/10 |
| Scan time (min: sec) | 4:02 |

Note. TSE = turbo spin echo, SPAIR = spectral attenuated inversion recovery, TE = echo time, TR = repetition time

**Table 2. Data Groups and Demographics**

|  | Patient (*n*=30)[1] | Patient (*n*=27)[2] | Healthy (*n*=10) |
|---|---|---|---|
| Average age (years) | 67.63 | 68.04 | 24.90 |
| Average BMI (kg/m$^2$) | 26.00 | 25.44 | 22.75 |
| Male [n (%)] | 9 (30.00) | 9 (30.00) | 5 (50.00) |
| Female [n (%)] | 21 (70.00) | 18 (60.00) | 5 (50.00) |
| K-L 0 [n (%)] | N/A | N/A | 10 (100%) |
| K-L 1 [n (%)] | 2 (6.67) | 2 (6.67) | N/A |
| K-L 2 [n (%)] | 13 (43.33) | 12 (40.00) | N/A |
| K-L 3 [n (%)] | 6 (20.00) | 6 (20.00) | N/A |
| K-L 4 [n (%)] | 9 (30.00) | 7 (23.33) | N/A |

Note. N/A = not applicable, [1]This group of patients was used in segmentation training and evaluation. [2]This group of patients was used in subregion parcellation experiments.

## Table 3. Automated Segmentation Performance

Segmentation performance

| Cartilage Compartment | DSC (patient, *n*=30) | ASSD (patient, *n*=30) | DSC (healthy, *n*=10) | ASSD (healthy, *n*=10) |
|---|---|---|---|---|
| Femoral cartilage | 0.78 | 0.45 | 0.88 | 0.21 |
| Medial tibial cartilage | 0.74 | 0.46 | 0.86 | 0.31 |
| Lateral tibial cartilage | 0.79 | 0.54 | 0.85 | 0.25 |

Note. DSC = Dice similarity coefficient (larger is better than smaller), ASSD = average symmetric surface distance (smaller is better than larger).

Regional mean spin-lattice relaxation time constant in the rotating frame ($T_{1\rho}$)

| Cartilage Compartment | Paired t-test (patient, *n*=30) | RMSD (*ms*, patient, *n*=30) | $CV_{RMSD}$ (%, patient, *n*=30) | Paired t-test (healthy, *n*=10) | RMSD (ms, healthy, *n*=10) | $CV_{RMSD}$ (%, healthy, *n*=10) |
|---|---|---|---|---|---|---|
| Femoral cartilage | *p*=0.88 | 0.28 | 0.65 | *p*=0.64 | 0.11 | 0.25 |
| Medial tibial cartilage | *p*=0.27 | 0.47 | 1.22 | *p*=0.71 | 0.32 | 0.82 |
| Lateral tibial cartilage | *p*=0.14 | 0.28 | 0.78 | *p*=0.29 | 0.34 | 0.96 |
| Average | N/A | 0.34 | 0.88 | N/A | 0.26 | 0.68 |

Note. RMSD = root-mean-square deviation (smaller is better than larger), $CV_{RMSD}$ = coefficient of variance of RMSD (smaller is better than larger).

**Table 4. Regional Mean Spin-Lattice Relaxation Time Constant ($T_{1\rho}$) Agreement in Cartilage Subregions**

| Subregion | Paired test (patient, $n$=27) | RMSD ($ms$, patient, $n$=27) | $CV_{RMSD}$ (%, patient, $n$=27) | Paired test (healthy, $n$=10) | RMSD (ms, healthy, $n$=10) | $CV_{RMSD}$ (%, healthy, $n$=10) |
|---|---|---|---|---|---|---|
| aMFC | $p$=0.63 | 0.71 | 1.51 | $p$=0.11 | 0.44 | 0.96 |
| ecMFC | $p$=0.34[#] | 0.79 | 2.11 | $p$=0.16 | 0.53 | 1.40 |
| ccMFC | $p$=0.06[#] | 1.73 | 4.58 | $p$=0.85 | 1.07 | 2.50 |
| icMFC | $p$=0.48[#] | 0.59 | 1.28 | $p$=0.66 | 0.27 | 0.57 |
| pMFC | $p$=0.92 | 0.60 | 1.30 | $p$=0.28 | 0.15 | 0.38 |
| aLFC | $p$=0.66 | 1.12 | 2.46 | **$p$=0.02*** | 0.37 | 0.78 |
| ecLFC | $p$=0.07[#] | 0.55 | 1.43 | $p$=0.45 | 0.67 | 1.70 |
| ccLFC | $p$=0.96 | 0.95 | 2.41 | $p$=0.71 | 1.24 | 3.01 |
| icLFC | $p$=0.43 | 1.18 | 2.62 | $p$=0.46 | 0.63 | 1.46 |
| pLFC | $p$=0.32 | 0.36 | 0.81 | $p$=0.65 | 0.20 | 0.47 |
| aMTC | $p$=0.67 | 0.73 | 1.82 | $p$=0.77[#] | 0.57 | 1.34 |
| eMTC | $p$=0.97[#] | 0.94 | 2.26 | $p$=0.12 | 0.52 | 1.23 |
| pMTC | $p$=0.70 | 0.50 | 1.31 | $p$=0.98 | 0.46 | 1.16 |
| iMTC | **$p$=0.02*** | 0.63 | 1.59 | $p$=0.66 | 0.40 | 0.98 |
| cMTC | $p$=0.28 | 1.09 | 3.09 | $p$=0.55 | 0.72 | 1.84 |
| aLTC | $p$=0.07 | 0.67 | 1.81 | **$p$<0.01*** | 0.45 | 1.27 |
| eLTC | **$p$=0.02*[#]** | 0.71 | 1.76 | $p$=0.23 | 0.95 | 2.30 |
| pLTC | $p$=0.13 | 0.64 | 1.61 | $p$=0.23 | 0.66 | 1.68 |
| iLTC | $p$=0.43 | 0.80 | 2.01 | $p$=0.66 | 0.51 | 1.47 |
| cLTC | **$p$=0.05*[#]** | 0.60 | 1.73 | $p$=0.32[#] | 0.34 | 1.05 |
| Average | N/A | 0.79 | 1.97 | N/A | 0.56 | 1.38 |

Note. **Bold text** and * means statistical significance. [#] denotes a *p*-value that was derived from a Wilcoxon test, as the corresponding normality test failed. Other paired tests were paired-sample *t*-tests. a = anterior, p = posterior, e = exterior, i = interior, c = central, ic = interior central, cc = central central, ec = exterior central, MFC = medial femoral cartilage, LFC = lateral femoral cartilage, MTC = medial tibial cartilage, LTC = lateral femoral cartilage.

# Supplementary Material

## A. Evaluation Metrics

### A.1 Segmentation Metrics

**Dice similarity coefficient (DSC)** is a common segmentation metric that measures the overlap between the prediction and ground truth. It can be formulated as

$$DSC = \frac{2TP}{2TP + FP + FN}$$

where *TP*, *FP*, and *FN* are true positive, false positive, and false negative, respectively. In our research, DSC was calculated for every pixel within the area of each class.

**Average symmetric surface distance (ASSD)** is a symmetric version of Average surface distance (ASD), which is another segmentation metric that measures the distance between the images. ASD can be expressed as follows:

$$ASD(X,Y) = \frac{\sum_{x \in X} \min_{y \in Y} d(x,y)}{|X|}$$

and ASSD can be expressed as:

$$ASSD(X,Y) = \frac{ASD(X,Y) + ASD(Y,X)}{2}$$

where $x \in X$ and $y \in Y$ are two sets of points, and *d(x, y)* is the Euclidean distance between a pair of points.

### A.2 Metrics for $T_{1\rho}$ Quantification

**Root mean squared deviation (RMSD)** measures the difference of the ground truth (reference) and predicted regional-mean $T_{1\rho}$ values. Giving them as $Q$ and $\hat{Q}$, and *n* samples, RMSD can be obtained by:

$$RMSD(\hat{Q}, Q) = \sqrt{\frac{1}{n}(\hat{Q} - Q)^2}$$

And the coefficient of variance of RMSD ($CV_{RMSD}$) can be acquired by:

$$CV_{RMSD}(\hat{Q}, Q) = \frac{RMSD(\hat{Q}, Q)}{\bar{Q}}$$

**Paired statistical tests on $T_{1\rho}$ quantifications.** We leveraged a paired t-test and a Wilcoxon test on the mean $T_{1\rho}$ values calculated on predicted and ground truth segmentation masks. We validated whether the mean $T_{1\rho}$ data from an individual ROI is normally distributed. A Wilcoxon test rather than a paired t-test for non-normally distributed ROIs will be applied.

# B. Additional Results

## B.1 Normality Test for Experiment 1 and 3

The normality tests were conducted using Shapiro-Wilk tests. The test results for **Experiment 1** and **Experiment 3** are shown in Table S1 and S2, respectively.

## B.2 Experiment 2

In the controlled study of **Experiment 2**, we chose two healthy volunteers, one for the left and one for the right knee, to test the robustness of parcellation against standardisation. Below are the additional results for the controlled rotation study of **Experiment 2**. Figure S1 gives the second half of the results (rotating in the opposite direction) for healthy volunteer V014. Additionally, Figure S2 gives a rotation experiment result on another healthy volunteer (V016).

**Table S1. Normality Test Statistics for Cartilage Masks**

| Cartilage Compartment | Ground truth (patient, $n$=30) | Predict (patient, $n$=30) | Ground truth (healthy, $n$=10) | Predict (healthy, $n$=10) |
|---|---|---|---|---|
| Femoral cartilage | $p$=0.37 | $p$=0.21 | $p$=0.45 | $p$=0.51 |
| Medial tibial cartilage | $p$=0.99 | $p$=0.24 | $p$=0.37 | $p$=0.35 |
| Lateral tibial cartilage | $p$=0.79 | $p$=0.62 | $p$=0.95 | $p$=0.34 |

**Table S2. Normality Test Statistics for Cartilage Subregions**

| Subregion | Reference (patient, $n$=27) | Predict (patient, $n$=27) | Reference (healthy, $n$=10) | Predict (healthy, $n$=10) |
|---|---|---|---|---|
| aMFC | $p$=0.27 | $p$=0.74 | $p$=0.73 | $p$=0.32 |
| ecMFC | **$p$<0.01*** | **$p$<0.01*** | $p$=0.33 | $p$=0.85 |
| ccMFC | **$p$<0.01*** | **$p$<0.01*** | $p$=0.42 | $p$=0.68 |
| icMFC | $p$=0.47 | **$p$=0.05*** | $p$=0.35 | $p$=0.68 |
| pMFC | $p$=0.71 | $p$=0.40 | $p$=0.97 | $p$=0.62 |
| aLFC | $p$=0.41 | **$p$<0.01*** | $p$=0.87 | $p$=0.66 |
| ecLFC | **$p$<0.01*** | $p$=0.96 | $p$=0.18 | $p$=0.54 |
| ccLFC | $p$=0.42 | $p$=0.10 | $p$=0.86 | $p$=0.09 |
| icLFC | $p$=0.34 | **$p$<0.01*** | $p$=0.68 | $p$=0.48 |
| pLFC | $p$=0.24 | $p$=0.84 | $p$=0.25 | $p$=0.14 |
| aMTC | $p$=0.9 | $p$=0.74 | **$p$=0.01*** | $p$=0.87 |
| eMTC | **$p$=0.01*** | $p$=0.13 | $p$=0.82 | $p$=0.04 |
| pMTC | $p$=0.62 | $p$=0.78 | $p$=0.71 | $p$=0.76 |
| iMTC | $p$=0.97 | $p$=0.63 | $p$=0.47 | $p$=0.12 |
| cMTC | $p$=0.07 | $p$=0.70 | $p$=0.14 | $p$=0.70 |
| aLTC | $p$=0.38 | $p$=0.12 | $p$=0.05 | $p$=0.29 |
| eLTC | **$p$=0.04*** | **$p$<0.01*** | $p$=0.44 | $p$=0.74 |
| pLTC | $p$=0.81 | $p$=0.19 | $p$=0.11 | $p$=0.78 |
| iLTC | $p$=0.64 | $p$=0.65 | $p$=0.11 | $p$=0.61 |
| cLTC | **$p$<0.01*** | **$p$=0.03*** | $p$=0.29 | **$p$=0.04*** |

Note. **Bold text** and * means statistical significance and the data samples in this subregion are not normally distributed. a = anterior, p = posterior, e = exterior, i = interior, c = central, ic = interior central, cc = central central, ec = exterior-central, MFC = medial femoral cartilage, LFC = lateral femoral cartilage, MTC = medial tibial cartilage, LTC = lateral femoral cartilage.

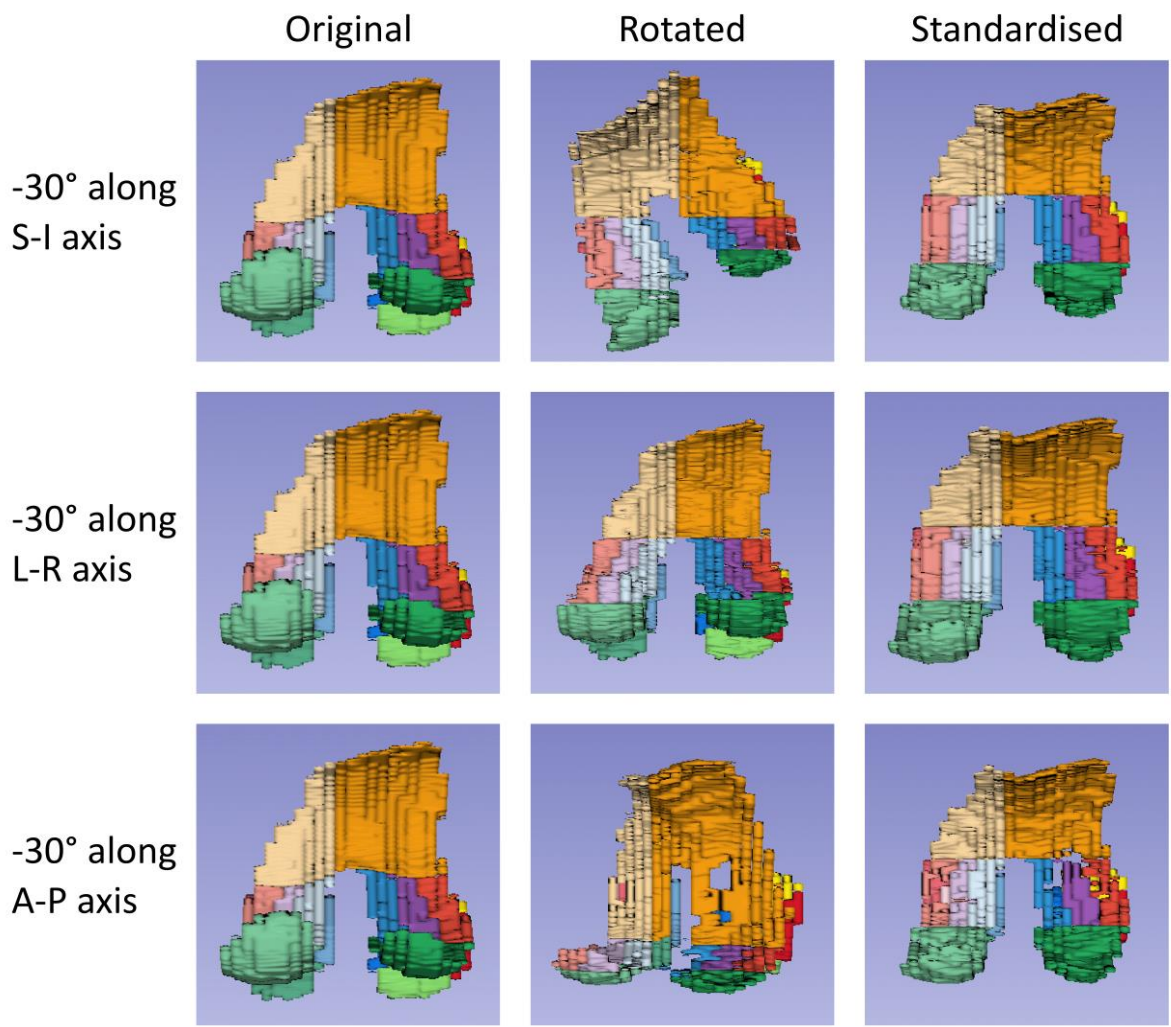

Figure S1. Negative direction of controlled rotation experiment on a healthy volunteer (V014, right knee, female, 22y, BMI = 20.76 kg/m$^2$). All shots of the 3D-rendered masks were taken from the unified camera position. Note: L = left, R = right, A = anterior, P = posterior, S = superior, I = inferior, BMI = body mass index.

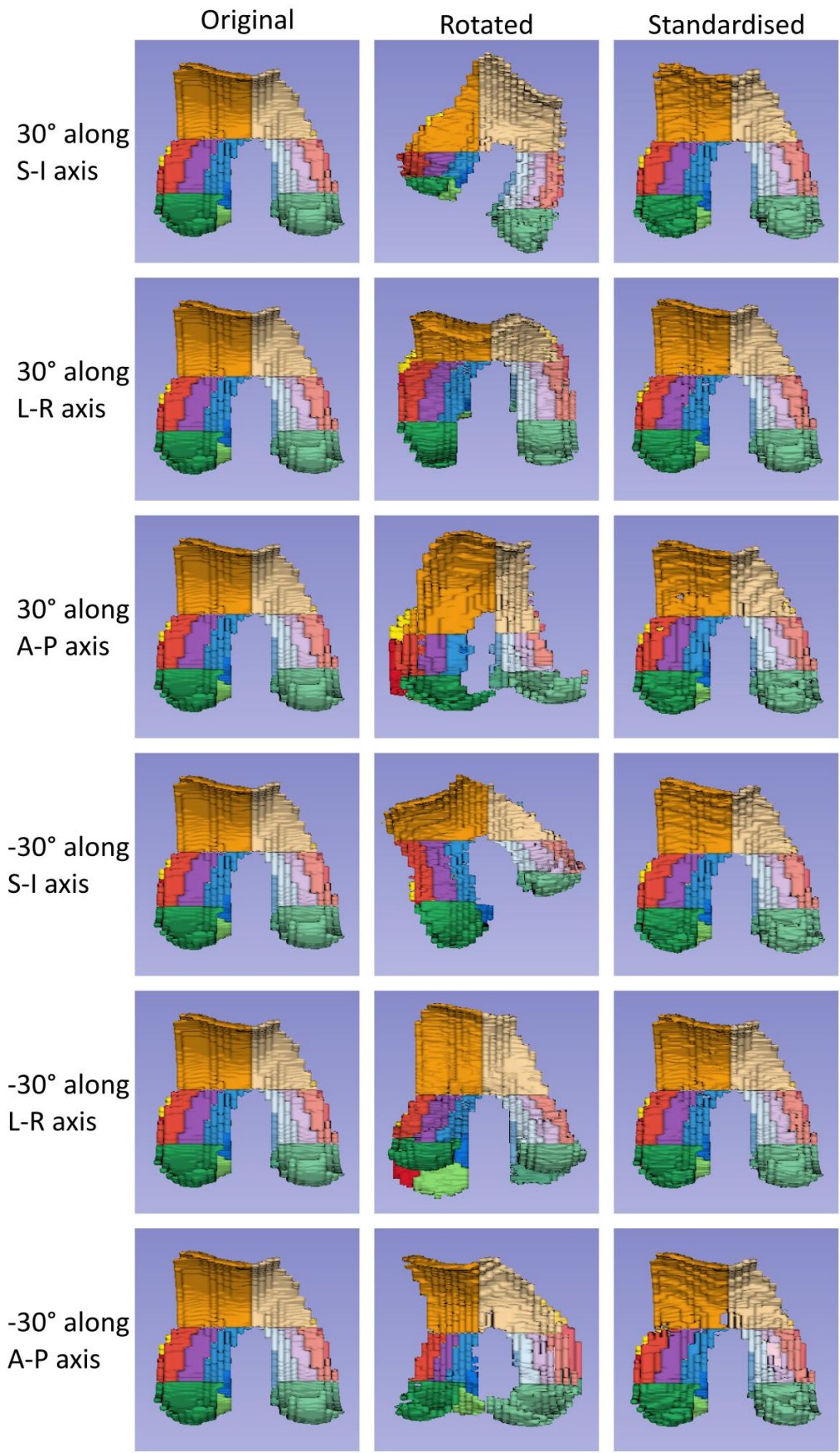

Figure S2. Controlled rotation experiment on a healthy volunteer (V016, left knee, male, 30y, BMI = 24.21 kg/m$^2$). All shots of the 3D-rendered masks were taken from the unified camera position. Note: L = left, R = right, A = anterior, P = posterior, S = superior, I = inferior, BMI = body mass index.